\documentclass[manuscript,screen]{acmart}

\AtBeginDocument{%
  \providecommand\BibTeX{{%
    \normalfont B\kern-0.5em{\scshape i\kern-0.25em b}\kern-0.8em\TeX}}}
\usepackage{colortbl}

\usepackage{multirow}
\begin{document}

\title{COVID-19 as Reflected in University President Bulk Email }






\author{Ruoyan Kong}
\email{kong0135@umn.edu}
\affiliation{%
  \institution{University of Minnesota - Twin Cities}
  \city{Twin Cities}
  \country{USA}
}

\author{Charles Chuankai Zhang}
\affiliation{%
  \institution{University of Minnesota - Twin Cities}
  \city{Twin Cities}
  \country{USA}
}

\author{Jin Kang}
\affiliation{%
  \institution{University of Minnesota - Twin Cities}
  \city{Twin Cities}
  \country{USA}
}

\author{Haiyi Zhu}
\email{haiyiz@cs.cmu.edu}
\affiliation{%
  \institution{Carnegie Mellon University}
  \city{Pittsburgh}
  \country{USA}
}

\author{Joseph A. Konstan}
\email{konstan@umn.edu}
\affiliation{%
  \institution{University of Minnesota - Twin Cities}
  \city{Twin Cities}
  \country{USA}
}


\begin{abstract}
E-mail ``Messages From the President'' to university students, staff, and faculty have long been used to keep campus communities aware of the latest policies, events, and news. But during the COVID-19 pandemic, as universities quickly closed facilities, sent students home, and canceled travel, these messages took on even greater importance.  We report on a content analysis of bulk emails from different universities' presidents to their students and employees before and in three stages of the pandemic. We find that these messages change as universities move towards and through closure. During the pandemic, 1) presidential bulk emails tend to be more informative, positive, clearer than before; 2) they tend to use more personal and collective language; 3) university presidents tend to mention more local political leaders and fewer other university leaders. Our results can inform research on digital crisis communication and may be useful for researchers interested in automatically identifying crisis situations from communication streams.  
\end{abstract}


\keywords{COVID-19, bulk email, organization crisis communication, remote work}

\maketitle

\vspace{-0.15in}\section{Introduction}
Universities around the world are experiencing a crisis --- they have closed facilities, sent students home, and canceled travel due to the COVID-19 pandemic. University presidents now need to conduct internal crisis communication (the communicative interaction within a private or public organization, before, during and after an organizational or societal crisis \cite{JOHANSEN2012270}), such as announcing pandemic-related news, remote-work policies, and health information.

``Message From the President'', which are the emails have long been used to keep campus communities aware of the latest policies, events, and news, now take on even greater importance because of the need of internal crisis communication \cite{kong2023socially}. For one side, now organization communications largely rely on telecommunication technologies such as emails, given that many conventional in-person communication channels are no longer available \cite{konstan2023challenge}. For another, they have to meet rapidly changing circumstances and conditions. These emails are called central bulk emails --- the internal emails that are sent to a large group of recipients from the central of an organization (e.g. university president's office).  


How do these central bulk emails change after the happening of the pandemic? How do the modes of designing, wording, and sending these emails different in a crisis? How do university presidents react to the pandemic in these emails? We would like to conduct a preliminary content analysis study around these questions because previous research, as we discuss below, do not focus on the usage of central bulk emails in internal crisis communication, but internal crisis management strategies \cite{LEE2019101832, SAMRA2019102038, PEDERSEN2020314,caillouet1996impression,huang2006crisis}, or the crisis communication with external public \cite{10.1145/3025453.3025627} \cite{10.1145/3173574.3173788} \cite{coombs2007protecting,sun2023less,sun2023interactive}.

First, we create a dataset of the central bulk emails of 8 United States universities, which are sent from their presidents'

\noindent offices to campus communities before and in three different stages of the pandemic. Second, we do a content analysis of these emails and compare emails between different stages in metrics around crisis communication strategies.


We find that in different stages of the pandemic, these central bulk emails have different patterns. During the pandemic, 1) presidential bulk emails tend to be more informative, positive, clearer; 2) they tend to use more personal and collective language; and 3) presidents tend to mention more local political leaders and fewer other university leaders. Our work could inform research on building systems to help organization leaders to select bulk-email communication strategies in crisis, or building models to automatically identify crisis situations from central bulk emails.





\vspace{-0.1in}\section{Related Work}
\subsection{Organizational Communication}
Organizational communication is defined as the collective and interactive process of generating and interpreting messages within organizations to achieve their purposes \cite{stohl1995organizational}. Previous studies found that information providers played an important role in organizational communication's effectiveness \cite{gwizdka2001supporting}. Randall \cite{SCHULER1979268} surveyed 382 employees of a manufacturing firm, found the vicious cycle phenomena --- if information producers failed to provide clear information, it was unlikely that the information receivers would continue to seek out information. Different stakeholders within organizations might also get different communication effectiveness. Danis et al. surveyed managers and non-managers, found that user's organizational role had an impact on their conceptualization and likely use of email \cite{danis2005managers}. 

Nowadays bulk email functions as task assignment and newsletter distribution tools in organizational communication \cite{Jackson:2003:UEI:859670.859673,kong2023towards}. However, organizational bulk email system was found to be inefficient and incomplete. Dev et al. investigated the user experience of unsubscribing from unwanted bulk emails and found frustration with the prevailing options for limiting access \cite{10.1145/3313831.3376165}. Dabbish and Cadiz collected emails from 6 employees' inboxes and found that an increase in the number of recipients of a message
caused a decrease in the probability of a message being
retained \cite{10.1145/765891.766073}. Park et al. surveyed 77 users and found the need of attention management of bulk emails on message notifications \cite{10.1145/3290605.3300604}. 

\vspace{-0.1in}\subsection{Organizational Crisis Communication}
Coombs \cite{coombs2009conceptualizing} defined crisis as the perception of an event that threatens expectancies of stakeholders, and defined crisis communication as the collection, processing, and
dissemination of information required to address a crisis situation. We concluded the crisis communication strategies identified by previous research as the following 6 categories:
\vspace{-0.07in}
\begin{enumerate}
    \item \textbf{Communicate information clearly in a timely manner:} Communicating information on time is the top priority. Fernandez and Shaw \cite{fernandez2020academic} in a research about academic leadership in COVID-19 proposed that the best practices for navigating
unpredictable challenges for leaders was communicating frequently and clearly.


    \item \textbf{Convey messages in a positive manner:} Leaders tend to communicate more positive emotion during crisis. 
    Johansson and B{\"a}ck \cite{johansson2017strategic} analyzed strategic leadership communication in a network and found that encouraging/facilitating ways of communication were employed during crisis. 
    

    \item \textbf{Create collectiveness inside community:} Collective effort was found to be extremely important in public health issues like COVID-19 \cite{ueda2020managing}. Marx \cite{marx2000ten} studied re-engaging the public in local schools and proposed that school should build a sense of community and ``we'' in this crisis.  
    
    \item \textbf{Emphasize leader's personal humanity:} Some leaders attempt to personalize their communication in crisis. Bligh et al. \cite{BLIGH2004211} examined the rhetorical content of President George W. Bush's public speeches before and after the terrorist attacks of September 11th and found that the language became more charismatic after the crisis.
    
    \item \textbf{Centralize leadership:} Strong leadership is often referred as the most appropriate response to a crisis \cite{grint2005problems}.
    
    \noindent Mulder et al. surveyed an organization in crisis and non-crisis situations; found agreement on ``leadership should not be shared'' during the crisis \cite{mulder1971organization}.
    
    \item \textbf{Refer to government:} It is expected that government would react and communicate with the public in crisis. Alexandra et al. \cite{10.1145/2675133.2675242} investigated tweets about human-induced disasters and found that in progressive disasters, tweets from government were the information that people tended to be more concerned with.
    
\end{enumerate}

\vspace{-0.15in}\subsection{Gap}
\vspace{-0.05in}Previous crisis communication research focused on internal crisis management strategies \cite{LEE2019101832, SAMRA2019102038, PEDERSEN2020314,caillouet1996impression,huang2006crisis}, or the crisis communication with external public, such as how should governments provide crisis information to the public \cite{10.1145/3025453.3025627}, how do the public perceive crisis information on social media \cite{10.1145/3173574.3173788}, or how do organizations protect their reputation in external stakeholders during crisis \cite{coombs2007protecting}. However, how do organization leaders (e.g., presidents) 
deliver information like policy update and community support through central bulk emails in internal crisis communication remains unknown. This problem is important because 1) organization communications largely rely on emails in the current remote-working environment; 2) bulk email is the major communication channel for organization leaders to announce rapidly changing policies/arrangements to general organization members  \cite{kong2020organizational}.



\vspace{-0.05in}\section{Research Question}
We are identifying differences between communication in crisis and normal periods. Specifically, we would like to study how do organizations' central bulk emails change during a crisis. We break down the changes into a set of dimensions around the crisis communication strategies we identified above: 1) quantity and clarity of emails; 2) positivity of email content; 3) use of collective/personal language; 4) reference to other school leaders/political leaders. The answers to these questions might illuminate the building of bulk email system to support organizations' internal crisis communication.

\vspace{-0.05in}\section{Data Collection}
We collect the emails that are sent from university presidents to students and employees, from the public websites of presidents' offices (for example, the president's messages of University of Tulsa is in https://utulsa.edu/archive/president-messages/). We collected 4 parts of data to represent different stages of the pandemic, each stage's length is 2 months; thus we analyze based on the same length of periods according to the definition below:

\begin{itemize}
    \item Non-Covid stage - Baseline stage 05/2019 - 06/2019, when the COVID-19 pandemic did not happen;specifically, we select this period as a corresponding period of 05 - 06/2020.                                                                                                              \item Covid stage -  EarlyCovid stage 01/2020 - 02/2020, when the pandemic happened mainly outside United States; universities started to notice the pandemic and reminded students and employees; campuses remained opened.

    \item Covid stage - CovidCrisis stage 03/2020 - 04/2020, when the pandemic happened across the United States; universities closed their campuses and started communicating remotely.                                     
    \item  Covid stage - CovidEra stage 05/2020 - 06/2020, when the pandemic became a ``new normal''; universities have closed their campuses for several months and were discussing the next steps.
\end{itemize}

 We select 8 universities whose presidents at least send monthly emails to their students and employees by a stratified sampling approach. Specifically, we balance the attributes of these universities --- 3 public/5 private universities; 4 small (size $\leq$ 10000)/4 large universities; 3 female/5 male presidents. There are also small/large private universities with female/male presidents, and large public universities with female/male presidents (see Table \ref{tab: school}). There are 198 emails in total sent from the 8 universities in these 4 stages. We will make this dataset public in this link: (blinded).

\vspace{-0.05in}\section{Operationalization}
In this section we discuss our operationalization of the patterns of crisis communication, corresponding to the crisis communication strategies we identified in Section 2.2. For each email, we calculate the following metrics. 

\begin{table}[!htbp]
\small{
\centering
\arrayrulecolor[rgb]{0.8,0.8,0.8}
\begin{tabular}{|c|c|c|c|c|} 
\hline
\textbf{University}             & \textbf{Type} & \textbf{Region} & \textbf{Size} & \multicolumn{1}{c!{\color{black}\vrule}}{\textbf{President Gender}}  \\ 
\arrayrulecolor{black}\hline
University of Tulsa         & Private       & South, US       & 3000          & Female                                                               \\ 
\arrayrulecolor[rgb]{0.8,0.8,0.8}\hline
Cornell University          & Private       & Northeast, US   & 15000         & Female                                                               \\ 
\hline
Tufts University            & Private       & Northeast, US   & 6000          & Male                                                                 \\ 
\hline
Columbia University         & Private       & Northeast, US   & 6000          & Male                                                                 \\ 
\hline
University of Minnesota     & Public        & Midwest, US     & 35000         & Female                                                               \\ 
\hline
University of Illinois      & Public        & Midwest, US     & 34000         & Male                                                                 \\ 
\hline
Wake Forest University      & Private       & Southeast, US   & 5000          & Male                                                                 \\ 
\hline
University of New Hampshire & Public        & Northeast, US   & 13000         & Male                                                                 \\
\hline
\end{tabular}
\arrayrulecolor{black}
\caption{Type/region/student size/president's sender of universities selected.}
\label{tab: school}}
\vspace{-0.33in}
\end{table}

\begin{enumerate}
    \item  \textbf{Confirm information clearly in a timely manner:} we measure it from the quantity of information sent and the clarity of email. The quantity of information includes the frequency of sending emails and the email length; higher frequency and longer emails means that information is updated on time. Clarity includes number of sections/links per email, and the title's uniqueness. Intuitively, more sections/links and distinct titles means to help recipients get the topics of the email before opening it/when reading it, compared to putting all content together without division/putting all the content of external links in the email/using the same titles for all emails.
    \begin{itemize}
        \item frequency: the number of emails sent per month by a university.
        \item length: the number of words in the email.
        \item \#section: the number of sections in an email; a section is a part of the email with a highlighted subtitle.
        \item \#link: the number of hyperlinks to other pages per 1000 words in the email.
        \item title\_unique: the uniqueness of the email's title, defined as the average 	inverse document frequency score (IDF) \cite{jones1972statistical} of all words in the title: $
            unique score(title) = AVG_{word \in title} IDF(word, D)$, where $
            IDF(word, D) = \log \frac{|D|}{|\{d\in D: word\in d\}|},
            D = \{titles \ of \ emails \in corresponding \ university \ \& \ stage\}$.
    \end{itemize}
    \item \textbf{Convey messages in a positive manner:} we calculate the positivity score of content and closing:
    \begin{itemize}
        \item content\_positivity: the positive score of the email content by TextBLob sentiment analysis \cite{loria2018textblob} (from -1: negative to 1: positive), whose sentiment analysis is performed on matching phrases in Tom de Smedt's pattern library \cite{de2012pattern}; as an example of what kinds of language lead to a high (or low) positive score, \textit{``Your shared sacrifice, professionalism and strength of character represent the best of TU.''}'s positive score is 1 and \textit{``The isolation of quarantine may feel especially painful during this season.''}'s positive score is -1.
        \item close\_positivity:  the positive score of the email closing \cite{loria2018textblob}; the closing is defined as the last paragraph of the body, also the paragraph before the ending, for example, \textit{``Be safe, enjoy your summer and we look forward to seeing you in the fall.''} (before \textit{``Sincerely,''}).
    \end{itemize}
    \item \textbf{Create collectiveness inside community:} we measure the use of collective language, as they often appear together with the encouragement of collective effort.
    \begin{itemize}
        \item \#community: as ``community'' and ``family'' are the words often referred to in the studies on conceptualizing community in the workplace \cite{doi:10.1111/j.1467-8543.2011.00852.x} \cite{voydanoff2001conceptualizing}, we count the times that ``community, communities, family, families'' are mentioned in this email per 1000 words.
        \item \#we: the number of times that ``we, us, our, ours'' are mentioned in this email per 1000 words.
    \end{itemize}
    \item \textbf{Emphasize leader's personal humanity:} we measure the subjectivity scores of email content and closing:
    \begin{itemize}
        \item content\_subjectivity: the subjectivity score (from 0:objective to 1: subjective) of the email content with TextBLob
        
        \noindent sentiment analysis \cite{loria2018textblob}. For example, \textit{``There’s an old English proverb that notes, `In a calm sea every is a pilot.' Although I can’t individually thank every one of you in person, as I would like to, I do want to express my
sincere appreciation ...''}'s score is 1 and \textit{``Details for the town hall meeting for staff at 8:30 am ...''}'s score is 0.
        \item close\_subjectivity: the subjectivity score of the email closing with TextBLob sentiment analysis \cite{loria2018textblob}.

    \end{itemize}
    \item \textbf{Centralize Leadership:} we measure the number of times the email refers to other university leaders:
    \begin{itemize}
        \item  \#other university leader: the number of times that the email mentions university leaders (except the president) in the content per 1000 words, including vice president, dean, chancellor, director, provost, chief. If someone is mentioned repeatedly we also count the times of mentioning. The signature of the email is not counted. We manually label each email by a group of 3 students.
    \end{itemize}
    \item \textbf{Refer to government:} we measure the number of times the email refers to local political leaders:
    \begin{itemize}
        \item \#local political leader: the number of times that the email mentions local political leader per 1000 words, like governor, mayor, county executive/mayor, leader of local public health department. If someone is mentioned repeatedly we count the times of mentioning. We manually label each email by a group of 3 students.
    \end{itemize}
\end{enumerate}
\vspace{-0.05in}Then for each stage, we calculate the average of each metric of all universities. Notice that frequency is the average of number of emails sent per month in each stage of all universities. We also calculate the standard deviation of mean value of each stage. As the average of metrics might have large differences between universities (e.g. most emails from University of New Hampshire have more than 5 sections and most emails from University of Illinois only have 1 or 2 sections), we account the university that an email belong to as a random effect in the standard deviation calculation.

\section{Content Analysis Result}
We are doing an initial content analysis here by drawing the tendencies of crisis communication patterns we identified, with the examples of quotes from our dataset, but not statistical tests.\vspace{-0.05in}
\begin{enumerate}
    \item \textbf{Quantity of information:} Subfigure (1) in Figure \ref{fig:sum} shows the average of metrics on quantity of information of each stage. We could observe that universities sent a larger amount of information in CovidCrisis stage compared to Baseline stage and PreCvoid stage. In Baseline stage and EarlyCovid stage, universities send around 1.8 emails/month with less than 500 words; in CovidCrisis stage, universities send over 5 emails/month with over 600 words. And in CovidEra stage, the communication frequency drops ($\leq$ 4 emails/month), which might be caused by that rapid policy changes have been notified in March and April. 

\begin{figure}[!htbp]
\centering
  \includegraphics[width=1\columnwidth]{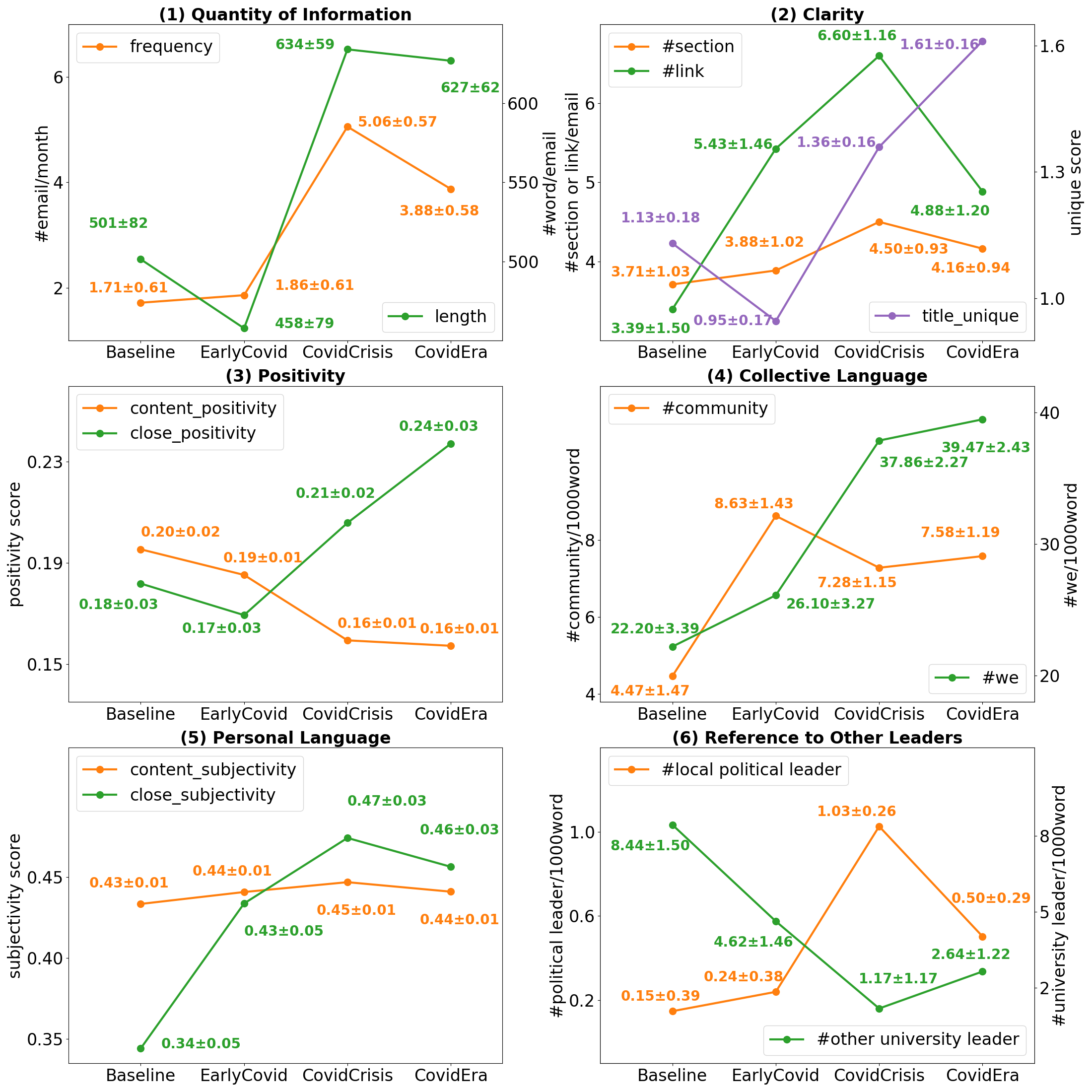}
  \caption{Average $\pm$ standard deviation of emails' metrics in each stage. Each subgraph includes the metrics that belong to a crisis communication pattern.}~\label{fig:sum}
  \vspace{-0.3in}
\end{figure}

\item \textbf{Clarity:} Subfigure (2) in Figure \ref{fig:sum} shows the average of metrics on clarity.
Titles are designed to be more unique in CovidCrisis stage (unique score = 1.36) and CovidEra stage (unique score = 1.61) compared to Baseline (unique score = 1.13) and EarlyCovid stages (unique score = 0.95). For example, most of University of New Hampshire's emails in Baseline and EarlyCovid stage have the same title \textit{``Update No. XX from President''}; and in CovidCrisis and CovidEra stage, the titles become more diverse, such as \textit{``UNH suspends all in-person classes for remainder of spring semester''} and \textit{``Managing the Financial Impact of COVID-19''}. Also, in Covid stages, emails have more sections ($\approx$ 4 sections/email) and more links ($\geq$ 4.8 links/email) compared to Non-Covid stage ($\leq$ 3.7 sections/email and $\leq$ 3.4 links/email). This shows that the emails in Covid stage tend to be clearer.

\item \textbf{Positivity:} Subfigure (3) in Figure \ref{fig:sum} shows the average of metrics on positivity of each stage. The figure shows the positivity of content in CovidCrisis and CovidEra stage decreases ($\approx 0.16$) while the positivity of closing paragraph increases ($\geq 0.21$) compared to Baseline and EarlyCovid stage (content\_positivity $\geq 0.19$, close\_positivity $\leq 0.18$ )  --- university presidents might close with a promising paragraph after telling some bad news. For example, the president of Cornell University closes with \textit{``I truly appreciate your ongoing attention to this unprecedented situation and have confidence that our community will make good choices, which will help to protect everyone.''} (positive score = 0.65) in the email \textit{``Cornell suspends classes; virtual instruction begins April 6''} (score = 0.09).


\item \textbf{Collective Language:} Subfigure (4) in Figure \ref{fig:sum} shows the average of metrics on collective language. University presidents mention more community words in Covid stage ($\geq$ 7 community words/1000 words) compared to Non-Covid stage (4.47 community words/1000 words) --- presidents remind employees/students that the community should work together to prevent the distribution of COVID-19 locally. For example, the president of Tufts University mentions the community concept 24 times/1000 words in an email in April with name \textit{``Helping local hospitals and communities respond to the COVID-19 surge''}, such as \textit{`` I have always been proud of the way the Tufts community rises to the occasion ...  have an abundance of resources to offer our community ... the Tufts community has stepped up in incredible ways, from students raising money ...''}. 

Also, presidents mention over 37 ``we'' words per 1000 words in CovidCrisis and CovidEra stage, while in Baseline and EarlyCovid stage this number is less than 26. For example, the president of Columbia University use ``we'' words 58 times/1000 words in an email which announced that the campus would be closed because of COVID-19, such as \textit{`` We must significantly reduce the number of students ...  We must take steps to reduce our research activities ...  we must continue to reduce ... we need to close various non-academic areas ...
''}.

\item \textbf{Personal Language:} Subfigure (5) in Figure \ref{fig:sum} shows the average of metrics on personal language of each stage. The subjectivity score of email content does not change significantly --- in CovidCrisis and CovidEra stage, most emails' content states policies and arrangements from an objective perspective. The subjectivity score of closing paragraph is larger in Covid stages ($\geq$ 0.43) compared to Baseline stage (0.34) --- presidents tend to show their personal feelings/humanities at the end of emails, for example, \textit{``Lastly, I’ll share one of my favorite Maya Angelou quotes, `No sun outlasts its sunset, but will rise again and bring the dawn.'...''} (score = 0.50, University of Minnesota).

\item \textbf{References to Other Leaders:} Subfigure (6) in Figure \ref{fig:sum} shows the average of metrics on references to other leaders of each stage. First, presidents mention fewer other university leaders in Covid stages ($\leq$ 5 mentions/1000 words) compared to Non-Covid stage (8.44 mentions/1000 words), especially in CovidCrisis stage (1.17 mentions/1000 words). This shows the centralization of leadership in university during crisis. 

Second, presidents mention more local political leaders in CovidCrisis stage (1.03 mentions/1000 words) than other stages ($\leq$ 0.5 mentions/1000 words) --- universities tend to refer to governments' guide/resources in crisis, such as \textit{``Today, Gov. J.B. Pritzker issued a stay-at-home order for residents of Illinois ... are sharing their expertise with the Governor’s Office ... Gov. Pritzker has praised the U of I System’s contributions''} (University of Illinois).

\end{enumerate}


\vspace{-0.08in}\section{Conclusion \& Discussion}
In this paper we study how does a crisis affect the dimensions of internal communication, specifically central bulk emails, as well as whether and how leaders change as we progress through post pandemic. We create a public dataset of universities' central bulk emails during different crisis stages of the COVID-19 pandemic, and conduct a content analysis of these central bulk emails to identify the trends and patterns of internal crisis communication in the pandemic.\vspace{-0.03in}
\begin{table}[!htbp]
\small{
\centering
\begin{tabular}{|p{5cm}|p{10cm}|} 
\hline
\textbf{Crisis Communication Strategy}                      & \textbf{Pattern During Covid Stage (Compared to Non-Covid Stage)}                                                                                       \\ 
\hline
Communicate information clearly in a timely manner. & Presidents (and their offices) send messages more frequently; these central bulk emails tend to contain more words/sections/links, with unique titles.  \\ 
\hline
Convey messages in a positive manner.                       & President's messages' closing are more positive while content are less positive.                                                                              \\ 
\hline
Create collectiveness inside community.                   & The uses of ``we'' words and ``community'' words increase.                                                                      \\ 
\hline
Emphasize leader’s personal humanity.                     & President's messages' closing are more subjective (while content maintains).                                                                            \\ 
\hline
Centralize leadership.                                  & Presidents (and their offices) mention fewer other university leaders in their messages.                                                                \\ 
\hline
Refer to government.                                        & Presidents (and their offices) mention more local political leaders in their messages.                                                                  \\
\hline
\end{tabular}
\caption{Matching between crisis communication strategies we identified and patterns of central bulk emails in the pandemic.}~\label{tab: match}}
\end{table}

\noindent\textbf{Limitations:} First, we are only doing content analysis to show the tendencies of central bulk emails in crisis, but not statistical tests. Second, our analysis only represents one way of operationalizing crisis communication patterns. We hope to collect data from more organizations to test whether the features of an organization (e.g. type/size/leader's gender of the organization) would influence its central bulk emails' patterns in crisis.

\vspace{0.06in}\noindent\textbf{Implications:} The changes of central bulk emails in crisis, to some extent, match crisis communication strategies, see Table \ref{tab: match}. This shows that university presidents and their offices are might applying crisis communication strategies in central bulk emails, 
intentionally or unintentionally.

Therefore first, for HCI researchers, we need to study the effectiveness of these crisis communication strategies in central bulk emails \cite{kong2021learning,aridor2022economics}. For example, when receiving central bulk emails in crisis, will recipients feel email overload because of larger quantity of information? Will recipients read and retain these information \cite{kong2022multi}? Will recipients feel connected/emotional supported? Will recipients prefer the communication style of personal language/strong leadership\cite{kong2021virtual,kang2020organizational}?

Second, for HCI developers, we could develop digital crisis communication system that could 1) automatically identify crisis situations from communication streams \cite{zhao2016group,he2023hiercat,kong2021nimblelearn}; 2) according to crisis situation, recommend crisis communication strategies and corresponding practices on central bulk emails to organization leaders\cite{zhang2022working}.

We hope that our results could inform research on digital crisis communication and be useful for researchers interested in automatically identifying crisis situations from communication streams.

\bibliographystyle{ACM-Reference-Format}
\bibliography{reference}


\end{document}